\documentclass[
reprint, 
superscriptaddress,
 amsmath,amssymb,
 aps, physrev,
 twocolumn,
prx,
]{revtex4-2}

\usepackage{comment}
\usepackage{xcolor}
\usepackage{graphicx}
\usepackage{dcolumn}
\usepackage{bm}
\usepackage{hyperref}
\usepackage{ulem}

\begin{document}

\preprint{APS/123-QED}

\title{Lattice quantum electrodynamics of a molecular emitter in a topological gap 
}%

\author{Clarisse Fournier}
\affiliation{Univ. Lille, CNRS, UMR 8523 -- PhLAM -- Physique des Lasers Atomes et Mol\'ecules, F-59000 Lille, France}

\author{Johannes Düreth}
\affiliation{Technische Physik, Wilhelm-Conrad-R\"ontgen-Research Center for Complex Material Systems and W\"urzburg-Dresden Cluster of Excellence ctd.qmat, University of W\"urzburg, Am Hubland, W\"urzburg,
97074, Bayern, W\"urzburg}

\author{Elena~Fanella}
\affiliation{National Institute of Optics (CNR-INO), c/o LENS via Nello Carrara 1, Sesto Fiorentino 50019, Italy}

\author{Hugo~Levy-Falk}
\affiliation{National Institute of Optics (CNR-INO), c/o LENS via Nello Carrara 1, Sesto Fiorentino 50019, Italy}

\author{Maja~Colautti}
\affiliation{National Institute of Optics (CNR-INO), c/o LENS via Nello Carrara 1, Sesto Fiorentino 50019, Italy}
\affiliation{European Laboratory for Non-Linear Spectroscopy (LENS), Via Nello Carrara 1, Sesto Fiorentino 50019, Italy}

\author{Anna~Beauvironnet}
\affiliation{Univ. Lille, CNRS, UMR 8523 -- PhLAM -- Physique des Lasers Atomes et Mol\'ecules, F-59000 Lille, France}

\author{Gauthier~Dekyndt}
\affiliation{Univ. Lille, CNRS, UMR 8523 -- PhLAM -- Physique des Lasers Atomes et Mol\'ecules, F-59000 Lille, France}

\author{Clément~Hainaut}
\affiliation{Univ. Lille, CNRS, UMR 8523 -- PhLAM -- Physique des Lasers Atomes et Mol\'ecules, F-59000 Lille, France}

\author{Tom\'a\v s~Neuman}
\affiliation{Institute of Physics, Czech Academy of Sciences, Cukrovarnick\'{a} 10, Prague, 16200, Czech Republic}

\author{Simon~Betzold}
\affiliation{Technische Physik, Wilhelm-Conrad-R\"ontgen-Research Center for Complex Material Systems and W\"urzburg-Dresden Cluster of Excellence ctd.qmat, University of W\"urzburg, Am Hubland, W\"urzburg,
97074, Bayern, W\"urzburg}

\author{Sven~H\"ofling}
\affiliation{Technische Physik, Wilhelm-Conrad-R\"ontgen-Research Center for Complex Material Systems and W\"urzburg-Dresden Cluster of Excellence ctd.qmat, University of W\"urzburg, Am Hubland, W\"urzburg,
97074, Bayern, W\"urzburg}

\author{Sebastian~Klembt}
\affiliation{Technische Physik, Wilhelm-Conrad-R\"ontgen-Research Center for Complex Material Systems and W\"urzburg-Dresden Cluster of Excellence ctd.qmat, University of W\"urzburg, Am Hubland, W\"urzburg,
97074, Bayern, W\"urzburg}

\author{Alejandro~González-Tudela}
\affiliation{Institute of Fundamental Physics IFF-CSIC, Calle Serrano 113b, 28006 Madrid, Spain}
\affiliation{Quantum Advanced Research Center (QuARC), CSIC, Calle Serrano 113b, 28006 Madrid, Spain}

\author{Costanza~Toninelli}
\affiliation{National Institute of Optics (CNR-INO), c/o LENS via Nello Carrara 1, Sesto Fiorentino 50019, Italy}
\affiliation{European Laboratory for Non-Linear Spectroscopy (LENS), Via Nello Carrara 1, Sesto Fiorentino 50019, Italy}

\author{Alberto~Amo}
\email{alberto.amo-garcia@univ-lille.fr}
\affiliation{Univ. Lille, CNRS, UMR 8523 -- PhLAM -- Physique des Lasers Atomes et Mol\'ecules, F-59000 Lille, France}

\date{\today}

\begin{abstract}
Engineering the photonic environment using lattices of coupled resonators, which we refer to as lattice quantum electrodynamics (QED), provides a route to control both the spontaneous emission of individual quantum emitters and the photon-mediated interactions between them. 
Here we introduce an optical lattice QED platform based on individual dibenzoterrylene (DBT) molecules embedded in anthracene crystals and coupled to lattices of open optical microcavities. 
This hybrid architecture benefits from narrow-linewidth molecular emitters, site-resolved optical access, engineered coupled-resonator bands, and compatibility with established molecular frequency-tuning techniques. 
As a proof-of-principle demonstration, we observe emitter-photon bound states formed when the optical transition of a single molecule is tuned to the band gap of a Su-Schrieffer-Heeger (SSH) cavity lattice. 
These in-gap states display directional localization and photon emission on a single sublattice, inherited from the vacancy-induced topological edge modes of the underlying SSH lattice.
Our results establish open-cavity lattices coupled to DBT molecules as a versatile architecture for engineering many-emitter quantum optical systems with controllable photon-mediated interactions.
\end{abstract}

\maketitle

\section{\label{sec:intro}Introduction}

Engineering the photonic environment of quantum emitters is a central strategy for controlling both their individual radiative properties and their collective interactions. In the paradigmatic setting of cavity quantum electrodynamics~\cite{kleppner_inhibited_1981,haroche_exploring_2006,ritsch13a}, a resonator reshapes the electromagnetic density of states into discrete modes, enabling Purcell enhancement or inhibition of spontaneous emission, coherent light--matter coupling, and photon-mediated interactions between emitters coupled to a common cavity field. 

A natural extension of this idea is to replace a single resonator by a structured photonic reservoir, such as an array of coupled cavities or a photonic crystal medium~\cite{chang2018,sheremet_waveguide_2023,gonzalez-tudela_lightmatter_2024}, which we refer to as lattice quantum electrodynamics. 
In these systems, the engineered photonic band structure provides control over the density of states, propagation, and spatial profile of the electromagnetic modes in a lattice environment, opening a route to tailor not only emission rates but also the range, directionality, and symmetry of emitter-emitter interactions over long distances. 
Depending on the lattice geometry, the spectrum may display conventional or topological band gaps, van Hove singularities, flat bands, or Dirac-like points, each giving rise to distinct individual and collective radiative dynamics~\cite{Gonzalez-Tudela2015b,douglas_quantum_2015,calajo_atom-field_2016,shi_bound_2016,bello_unconventional_2019,leonforte_vacancy-like_2021,Vega2021a,vega_topological_2025,DeBernardis2021,gonzalez-tudela_quantum_2017,gonzalez-tudela_markovian_2017,gonzalez-tudela_anisotropic_2019,Navarro2024,Dibenedetto2025,gonzalez-tudela_exotic_2018,Perczel2020a,Navarro-Baron2021a,Bello2022,Tecer2024,Calajo2025,Tecer2026,Pinto2025}. 

When an emitter is resonant with propagating band modes, its emission can be enhanced, becomes highly directional, or acquires a strongly non-Markovian character, enabling collective phenomena such as superradiance and subradiance~\cite{gonzalez-tudela_markovian_2017,gonzalez-tudela_quantum_2017}. 
In the opposite regime, when the emitter transition lies inside a photonic band gap, emission into propagating modes is suppressed and light--matter coupling produces an emitter-photon bound state localized around the emitter~\cite{bykov75a,john90a,kurizki_two-atom_1990}. The spatial extent and symmetry of these bound states are controlled by the band structure of the lattice, providing a mechanism to engineer photon-mediated interactions with tunable range, directionality, and topology~\cite{Gonzalez-Tudela2015b,douglas_quantum_2015,bello_unconventional_2019,leonforte_vacancy-like_2021,Vega2021a,vega_topological_2025,DeBernardis2021,Dibenedetto2025,calajo_atom-field_2016,shi_bound_2016}, and opening routes toward many-body states with non-trivial spatial order of interest in quantum simulation~\cite{Gonzalez-Tudela2015b,douglas_quantum_2015,Bello2022,Tecer2024,Calajo2025,Tecer2026}.

\begin{figure*}[t!]
\includegraphics[width=\textwidth]{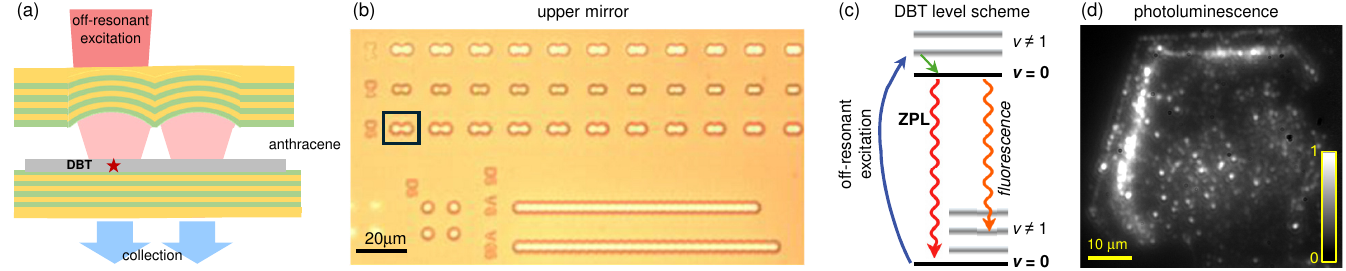}
\caption{\label{fig:cavity} 
The open cavity - DBT system. (a) Sketch of the open cavity with two coupled hemispheric resonators. (b) Optical microscope image of a region of the upper mirror displaying individual hemispheres, dimers and lattices of coupled hemispheres. (c) Simplified energy level scheme of a DBT molecule. $v=0$ indicates the fundamental vibrational state of the ground and first excited electronic state. ZPL is the zero-phonon line emission. (d) Photoluminescence image of the anthracene crystal with DBT molecules used in the experiments under wide field laser illumination at 764.150~nm, corresponding to the excitation from the full ground state to the first vibrational band $v=1$ of the first excited electronic state. The intensity scale has been saturated for visualization purposes.}
\end{figure*}

These perspectives have motivated experimental implementations of lattice quantum electrodynamics across several platforms. In the optical domain, natural atoms coupled to nanophotonic structures provide excellent coherence and intrinsic reproducibility~\cite{goban13a,thompson13a,goban15a,hood16a,Beguin2020a,Zhou2023,kim_trapping_2019,Samutpraphoot2020,Dordevic2021,Menon2024}, but scaling these systems requires trapping and positioning many atoms with subwavelength precision near structured photonic modes while maintaining stable light--matter coupling. Solid-state emitters, such as quantum dots~\cite{dietrich_gaas_2016, jurkat_single-photon_2023, chu_independent_2023,hallett_controlling_2026} and color centers~\cite{evans18a,Rugar2020,Rugar2021,lukin_two-emitter_2023} integrated with photonic-crystal structures, offer strong confinement and chip-scale compatibility, but deterministic emitter placement, spectral inhomogeneity, fabrication disorder, and independent frequency control remain major obstacles for realizing many-emitter devices. In parallel, superconducting qubits coupled to microwave resonator arrays or waveguides~\cite{liu17a,Sundaresan2019,Mirhosseini2019,Scigliuzzo2022,Kim2020b,Zhang2022,Owens2022,jouanny_superstrong_2025} and matter-wave implementations~\cite{krinner18a,Kwon2022FormationLattice,Kim2024a} have reached a high level of control, enabling observations of non-Markovian decay~\cite{krinner18a}, atom-photon bound states~\cite{liu17a,Kim2020b}, and tunable-range interactions~\cite{Scigliuzzo2022}. However, both implementations present challenges for harnessing optical quantum technologies: circuit QED platforms operate in a range far from visible and telecom wavelenghts; in matter-wave platforms, photons are only emulated, not actually realized.
An outstanding challenge is therefore to devise an optical lattice-QED platform that combines engineered photonic bands with spatially addressable, narrow-linewidth, and frequency-tunable quantum emitters.

In this work, we introduce an optical lattice-QED platform that couples lattices of open optical microcavities to individual dibenzoterrylene (DBT) molecules embedded in anthracene crystals, see Fig.~\ref{fig:cavity}. 
This hybrid architecture separates the engineering of the photonic lattice from the choice and control of the quantum emitter. On the emitter side, DBT molecules in anthracene crystals provide high-purity single-photon emission around the zero-phonon line in the near infrared, lifetime-limited linewidths at temperatures below $\sim 2$~K, negligible spectral diffusion, a limited residual inhomogeneous broadening of about 1--2~meV $(250$--$500~\mathrm{GHz})$~\cite{toninelli_single_2021}, and compatibility with optical and electrical frequency-tuning techniques~\cite{colautti_laser-induced_2020,duquennoy_enhanced_2024}. 
The combination of all these features has allowed for the observation of two-photon interference in the emission of distinct molecules~\cite{Duquennoy2022, huang_-chip_2025}.
On the photonic side, open microcavities provide site-resolved optical access and a versatile route for engineering coupled-resonator lattices~\cite{flatten_spectral_2016, dufferwiel_tunable_2015, dusel_room_2020}. In the single-resonator limit, their coupling to DBT molecules has already enabled single-molecule strong coupling~\cite{wang_turning_2019}, single-photon nonlinearities~\cite{pscherer_single-molecule_2021}, and cavity-mediated hybridization between two molecules~\cite{nobakht_hybridization_2025}. 

As a proof-of-principle demonstration of this lattice platform, here we study a single molecule coupled to photonic dimers and to one-dimensional Su-Schrieffer-Heeger (SSH) lattices. 
When the molecular transition is aligned with the photonic gap, we observe \textit{vacancy-like dressed states}~\cite{bello_unconventional_2019, leonforte_vacancy-like_2021}.
This is a special class of emitter-photon bound states characterized by a photonic distribution that corresponds to an eigenstate of the lattice of coupled photonic cavities with a vacancy at the emitter’s site~\cite{bello_unconventional_2019, leonforte_vacancy-like_2021, kim_quantum_2021}.
In the SSH lattice, they inherit the sublattice polarization and directional localization characteristic topological edge modes.
Beyond this demonstration, the platform opens a route toward many-emitter lattice QED with optically addressable molecules, engineered band structures, and controllable photon-mediated interactions.

\section{Open-Cavity Platform for Lattice QED with molecules}
To engineer lattices of coupled photonic resonators, we use an open cavity sketched in Fig.~\ref{fig:cavity}(a). It is made of two distributed Bragg mirrors (DBR) of 11 (bottom) and 12 (top) pairs of $\lambda/4$ TiO$_2$/SiO$_2$ layers on silica substrates, 500~$\mu$m thick.
The design wavelength is 782~nm, corresponding to the zero-phonon line of DBT in anthracene crystals.
Before the DBR deposition, the top mirror is etched using focused ion beam milling in the form of individual and coupled hemispheres~\cite{dusel_room_2020} with a diameter of 5~$\mu$m and different depth semi-axes of 330~nm and 480~nm, resulting in radii of curvature at the center of the hemispheres of 19~$\mu$m and 13~$\mu$m, respectively, see Fig.~\ref{fig:cavity}(b).
The bottom mirror is flat and has an extra layer of 30~nm of SiO$_2$ on top of the DBR to position the anthracene crystal with DBT molecules at an antinode of the confined field and improve the light-matter coupling.
To facilitate the alignment, the working region of both mirrors is located on top of a square mesa of lateral size of $400~\mu$m and a height of 50~$\mu$m, etched on the silica substrates.
The two mirrors are mounted on piezoelectric actuators that allow alignment at 5~K in a closed-cycle He cryostat in transmission geometry (see Appendix~\ref{app:set-up} for a detailed description of the set-up).

To characterize the intrinsic optical losses of the cavity, we measure the photon lifetime in individual hemispheric resonators prior to the insertion of the anthracene crystals. 
For the longitudinal cavity mode with index $q=9$, which is used throughout the experiments reported below, we obtain a photon lifetime of 50 ps from streak-camera measurements.
This value corresponds to a quality factor of the empty cavity of 120~000 and a finesse of 13~300.

On top of the bottom mesa, we place anthracene crystals with embedded DBT molecules using a fiber tip to transfer them from the sublimation growth chamber to the substrates~\cite{nobakht_hybridization_2025}.
Their typical lateral size is a few tens of microns, and their estimated thickness is of the order of a few hundreds of nanometers.
We then spin coat a thin layer of polyvinyl alcohol to protect them from sublimation.
The dipole of the DBT molecules is aligned parallel to the plane of the anthracene crystal~\cite{nicolet_single_2007}.
Figure~\ref{fig:cavity}(d) displays the photoluminescence of the crystal at 5~K, in the absence of the top mirror, under off-resonant wide field illumination at 764.150~nm.
This off-resonant wavelength efficiently excites the vibrational band of the first singlet electronic excited state of individual DBT molecules (see simplified level diagram in Fig.~\ref{fig:cavity}(c)).
After a fast non-radiative relaxation to the lowest level of the vibrational manifold (green arrow in Fig.~\ref{fig:cavity}(c)), molecules decay to the electronic ground state emitting a single photon (red transition in Fig.~\ref{fig:cavity}(c)).
About 1/3 of theses photons are emitted to the full ground state of the molecule at a wavelength of $\sim 782$~nm (the zero-phonon line).
The rest is emitted in combination with molecular vibrational excitations and/or phononic excitations of the anthracene matrix at longer wavelengths~\cite{pazzagli_self-assembled_2018, toninelli_single_2021}.

The photoluminescence in Fig.~\ref{fig:cavity}(d) is recorded in reflection geometry using a long-pass filter that suppresses the excitation wavelength.
A number of individual molecules is observed all over the crystal.
Their actual density is much higher than what is observed in Fig.~\ref{fig:cavity}(d).
However, the employed excitation wavelength only excites efficiently the small subset whose first vibrational band is resonant with the laser at 764.150~nm.
Local crystal fields shift this band for other subsets of molecules.
In the following, we will focus on the zero-phonon line emission of an individual molecule in the cavity under such off-resonant excitation. 

\begin{figure}[t]
\includegraphics[width=\columnwidth]{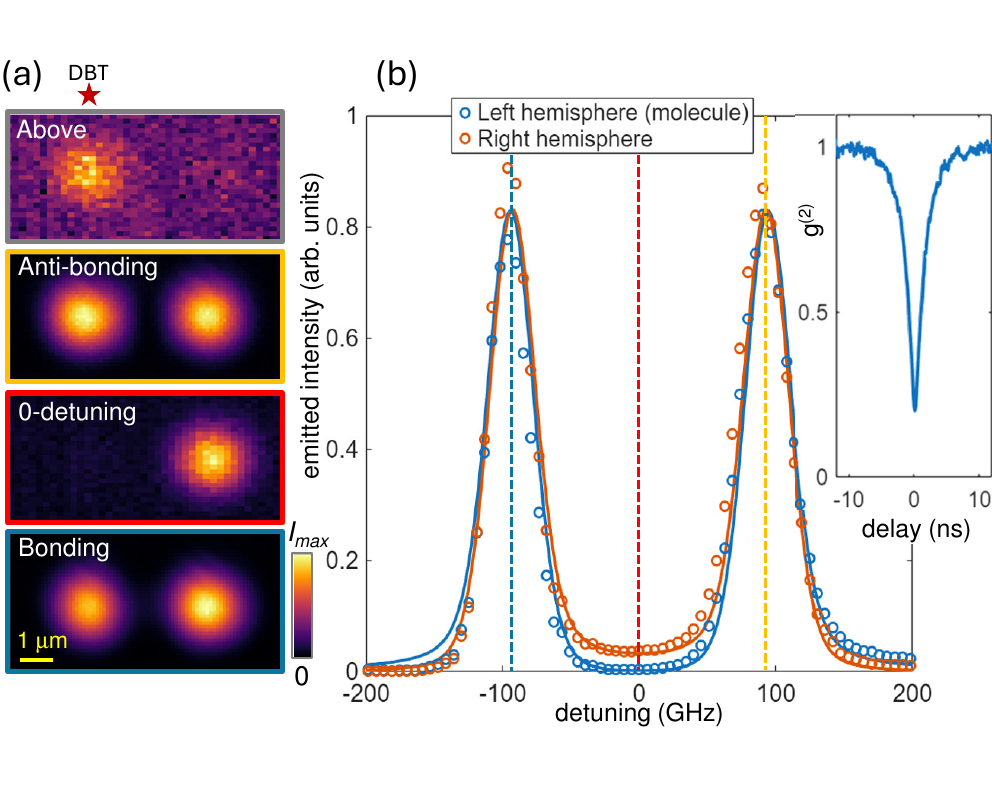}
\caption{\label{fig:dimer} 
DBT in a dimer of resonators. (a) Spatially resolved photoluminescence of a DBT molecule in a photonic dimer under off-resonant excitation at 765.267~nm at different emitter-cavity detunings $\Delta$: the color of the frame corresponds to the detunings marked in dashed lines in (b). For the top panel $\Delta=273$~GHz. 
The molecule is located close to the center of the left hemisphere. 
(b) Emitted intensity from the left (red dots) and right (blue dots) hemispheres as a function of the emitter-cavity detuning.
Solid lines represent the expected transmission including the effect of mechanical vibrations of the cavity length. The inset displays the measured $g^{(2)}$ when the molecule transition is in resonance with the bonding mode.}
\end{figure}

\section{Vacancy-like dressed state in a photonic dimer}

An insightful study of the formation of vacancy-like dressed states can be done in the simple configuration of a dimer of coupled hemispheres sketched in Fig.~\ref{fig:cavity}(a).
The tunneling of photons between the hemispheric resonators results in the formation of extended bonding and anti-bonding photonic modes arising from the coupling of the lowest energy transverse modes of the individual hemispheres with identical cavity length.
For a center-to-center separation between hemispheres of 3.55~$\mu$m and a semi-axis depth of 330~nm, at longitudinal mode index $q=9$, the energy splitting of these modes is $2 \times J =185.2$~GHz (765.9~$\mu$eV), with $J$ being the hopping energy.
Note that the birefringence of the anthracene crystal splits by 8~meV the photonic modes in linear polarization aligned along the fast and the slow axes of the crystal.
The dipole of the DBT molecules is oriented along the slow axis, and their emission is linearly polarized parallel to it.

Figure~\ref{fig:dimer}(a) displays spatial images of the measured photoluminescence under off-resonant excitation of a single molecule placed close to the center of the left hemisphere.
Each panel shows the photoluminescence at different emitter-cavity detunings $\Delta=E_{zpl}-E_0$, where $E_{zpl}$ is the zero-phonon line energy and $E_0$ the cavity resonance, which we vary by opening and closing the cavity with a piezoactuator.
For detunings of $\Delta = -92.6$~GHz and $\Delta = +92.6$~GHz, the molecular transition is in resonance with the bonding (bottom panel - blue frame) and antibonding (second panel from the top - yellow frame) modes, respectively.
In both cases, the spatial distribution of the emission is approximately equally distributed on both hemispheres.
Photon correlations measurements when the bonding mode is tuned to the emitter's frequency show a value of $g^{(2)}=0.2$ at zero time delay (see inset of Fig.~\ref{fig:dimer}(b) measured using a standard Hanbury Brown and Twiss (HBT) set-up), indicating that the emission originates from a single molecule. This $g^{(2)}$ measurement was done collecting the light emitted by the whole bonding mode from the two hemispheres.

At high positive detunings ($\Delta=273$~GHz, top panel of Fig.~\ref{fig:dimer}(a)), well above the antibonding mode, the emission is fully localized at the hemisphere where the molecule is located, resulting in a localized state similar to emitter-photon bound states in one-dimensional photonic lattices when the emitter energy is above or below the photonic band~\cite{bykov75a,john90a,kurizki_two-atom_1990}.

At $\Delta=0$, when the emitter transition energy resonates with the bare photon energy of each individual hemisphere, the nature of the emitter–photon localized state changes radically.
In this regime, the photons are localised at the hemisphere opposite to the one hosting the emitter (see the red-framed panel in Fig.~\ref{fig:dimer}(a)).
The emitted intensity measured at the center of each hemisphere, plotted as a function of the cavity–emitter detuning in Fig.~\ref{fig:dimer}(b) (see Appendix~\ref{app:data_analysis} for details), clearly reveals the localized character of the emission: at zero detuning, the emission from the left hemisphere drops to values near zero.

The localized photon states measured in the experiment can be modeled using a dissipative cavity system fed by a monochromatic source (the zero-phonon line emission of the molecule under off-resonant pumping)~\cite{gonzalez-tudela_connecting_2022}. 
In the tight binding limit and in the rotating frame of the emitter transition frequency, the coupled hemispheres-molecule system can be described using the following equation:
\begin{align}
i\hbar \frac{d\psi_{m}}{dt} &= (\Delta - i \frac{\kappa}{2})\psi_{m} + J \sum_{\langle n \rangle} \psi_{n}+i\sqrt{\frac{\kappa}{2}} F_m,
\label{eq:classical}
\end{align}
where $\psi_{m}$ is the photon field at hemisphere $m$, 
$\kappa$ is the cavity loss, $J$ is the photon hopping energy between nearest-neighbor sites.
For convenience, in this article we express energies in frequency units (GHz).
$F_m$ is the field emitted by the molecule located at site $m$, which is proportional to the dipole of the molecule and its coupling to the field in the cavity. In an empty site, $F_m=0$. 
For the dimer we are considering, $m = 1,2$, and the emitter is located on the left hemisphere $m=1$.
In this framework, the cancellation of the emission at the site of the emitter in the steady state arises from destructive interference of the emitted field which, at exactly $\Delta=0$, picks opposite signs from the red detuned bonding resonance and from the blue detuned antibonding resonance. 

Solid lines in Fig.~\ref{fig:dimer}(b) display the calculated photon intensity from Eq.~\eqref{eq:classical} including the effects of vibrations of the cavity caused by the closed-cycle cryostat and assuming no spectral diffusion of the emitter.
Mechanical vibrations induce a normal distribution of cavity lengths and a corresponding distribution of cavity energies that are averaged over the photoluminescence integration time.
A fit to the model produces $\kappa = 11~$GHz and a vibration-induced cavity energy variance of $\sigma = 34$~GHz (see Appendix~\ref{app:convolution} for details).
The intrinsic cavity loss $\kappa$  extracted from the fit is almost 4 times larger than that measured in an empty cavity, indicating that inhomogeneities associated with the anthracene–PVA deposition might degrade the quality factor.

A similar localized state at $\Delta=0$ in a lossless dimer has been predicted in Ref.~\cite{leonforte_vacancy-like_2021}.
It was interpreted as a vacancy-like dressed state, a special class of emitter-photon bound states characterized by the fact that its photonic part corresponds to an eigenstate of the photonic chain of resonators with a vacancy at the emitter's site.
In this case, a vacancy on the left hemisphere results in an eigenstate fully localized on the right hemisphere at photon energy $E_0$, the energy of the bare hemisphere.
The origin of these states can be traced down to the mirror effect of a two-level system in resonance with an electromagnetic field, which has been extensively studied in the context of atom-photon interactions~\cite{shen_coherent_2005, chang_single-photon_2007,zhou_controllable_2008, zhou_quantum_2008}.

For completeness, a dimer with different bare energies in the individual hemispheres is studied in Appendix~\ref{app:dimer_imbalance}.
Interestingly, this situation can be readily accessed by tilting the upper mirror with the use of piezoactuators, resulting in slightly different cavity lengths for each hemisphere~\cite{dufferwiel_tunable_2015}.
The bound state is still present but at a different detuning.

\begin{figure}[t]
\includegraphics[width=\columnwidth]{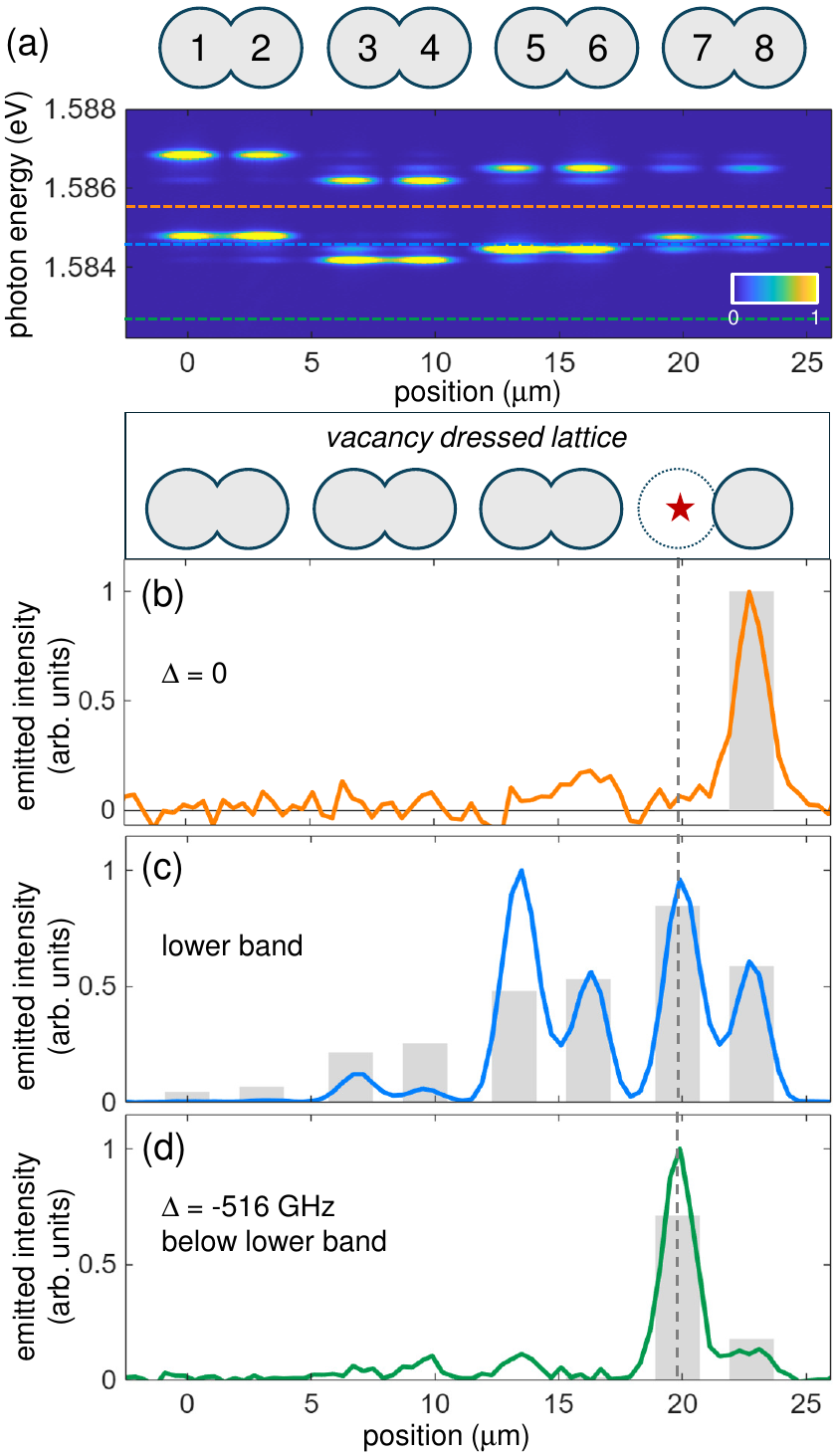}
\caption{\label{fig:SSH} 
Vacancy-like bound state in an SSH lattice. (a) Transmission spectrum across the center of the lattice sketched on the top using a broadband lamp.
(b)-(d) Solid lines display the measured photoluminescence at the transition energy of a molecule located at site $m=7$ (indicated by a red star and vertical dashes lines) for different emitter-hemisphere detunings.
The grey bars are the steady-state solutions of Eq.~\eqref{eq:classical} at the corresponding detunings.}
\end{figure}

\section{Vacancy-like dressed states in an SSH lattice}

Vacancy-like dressed states can be used to produce photonic topological edge states in the bulk of an SSH lattice.
This idea was proposed in Ref.~\cite{bello_unconventional_2019, leonforte_vacancy-like_2021} and observed in the microwave regime using superconducting cavities and transmon qubits in Ref.~\cite{Kim2020b}.
Figure~\ref{fig:SSH}(a) displays the schematics of an SSH lattice of eight coupled hemispheres with staggered center-to-center distances (3 and 3.6~$\mu$m) and, therefore, staggered hopping energies.
For this lattice, the semi-axis depth of the hemispheres is 480~nm.
Transmission spectra under white light illumination reveal two bands of modes separated by a gap (Fig.~\ref{fig:SSH}(a)).
Local disorder of the on-site energies at each hemisphere is apparent in the form of energy variations at different positions, with a standard deviation of 50~GHz. This value is at least six times smaller than the average central gap. 
Therefore, we will neglect them in the upcoming study.
The origin of the disorder is likely the thickness variations of the polyvinyl alcohol layer.
From the measured bands, we extract an average tight binding intracell hopping $J=234$~GHz and an intercell hopping $J'=48$~GHz.

In the following we show three different situations in which vacancy-like dressed states appear at $\Delta = 0$, in the middle of the central gap.
First, we place a single molecule at site $m=7$, the previous to the last site of the lattice.
Figure~\ref{fig:SSH}(d) shows the emission at a negative detuning $\Delta=-516$~GHz, when the transition frequency of the molecule is well below the lower band. 
The photoluminescence exits the cavity structure through the low transmission tail of the lower band. The emission is concentrated at site $m=7$ and, more weakly, at site $m=8$.
We use this measurement to check the position of the molecule in the lattice.
At the energy of the lower band, Fig.~\ref{fig:SSH}(c), the photoluminescence is distributed over three unit cells.
At $\Delta =0$, in the middle of the gap, a vacancy-like dressed state is formed with the emission localised at the last site $m=8$, with negligible emission from any other site including the one in which the molecule is located ($m=7$), see Fig.~\ref{fig:SSH}(b).
The gray bars in Fig.~\ref{fig:SSH}(b)-(d) display simulations of Eq.~\eqref{eq:classical} for the corresponding detunings.
The agreement is good in the gaps.
However, the band modes are quite sensitive to disorder and the experimental profile deviates from the tight-binding model without disorder.
We will observe the same features in the two cases discussed below.

The geometry of the emission profile of the vacancy-like dressed state can be easily understood from an equivalent photonic lattice with a vacancy at the location of the emitter, which is sketched at the top of Fig.~\ref{fig:SSH}(b).
To the left of the vacancy, the dressed SSH lattice ends in a topologically trivial termination (strong intracell hopping), and at the energy corresponding to $\Delta=0$ it does not have any eigenmode.
To the right of the vacancy, a single hemisphere is found.
The photon energy of its eigenmode corresponds precisely to $\Delta=0$, and an emitter-photon bound state is formed and located at that site ($m=8$).

\begin{figure}[t]
\includegraphics[width=\columnwidth]{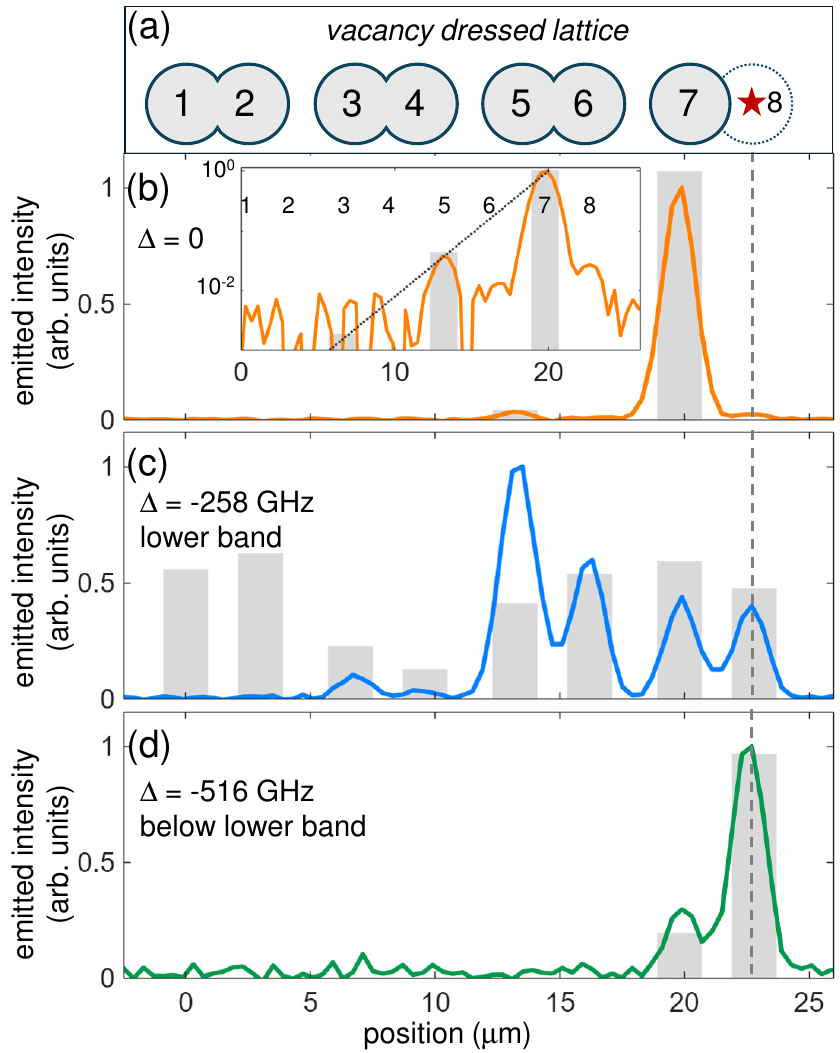}
\caption{\label{fig:chiral_left} 
Directional vacancy-like dressed state with molecule in hemisphere $m=8$. (a) Scheme of the vacancy dressed photonic lattice at the energy of the emitter-photon bound state at $\Delta=0$.
(b)-(d) Solid lines display the measured photoluminescence at the transition energy of a molecule located at site $m=8$ (indicated by a red star and vertical dashes lines) for different emitter-hemisphere detunings.
The grey bars are the steady-state solutions of Eq.~\eqref{eq:classical} at the corresponding detunings.
The inset of (b) displays the data in semi-log scale.
The dotted line shows the expected exponential decay of a topological edge state.}
\end{figure}

Directional vacancy-like dressed states are possible when the emitter is placed at other locations.
Figure~\ref{fig:chiral_left} displays the photoluminescence at different cavity detunings for an emitter at the edge site $m=8$.
As in the case discussed above, the emission at very negative detunings, when the transition energy of the emitter is below the gap, indicates the precise position of the emitter (Fig.~\ref{fig:chiral_left}(d)).
Figure~\ref{fig:chiral_left}(b) shows the emission of the vacancy-like dressed state at $\Delta=0$.
With the emitter at site $m=8$, the vacancy dressed lattice in Fig.~\ref{fig:chiral_left}(b) ends in a topologically non-trivial termination, and the photonic part of the emitter-photon bound state takes the form of a topological edge state.
It decays exponentially away from the vacancy and it is only present in the sublattice of odd hemispheres (intensity at sites $m=7$ and $m=5$ in the experiment). 
The dotted lines in the inset show the expected decay computed from the well-known analytics of topological edge states in SSH ($|\psi|^2\sim e^{\ln(J'/J)(\tilde m-\tilde m_0)}$ with $\tilde m$ and $\tilde m_0$ the unit cell number and edge unit cell, respectively), using the values of $J$ and $J'$ measured in this lattice~\cite{Delplace2011, Asboth2015}.
The observed distribution is reproduced by Eq.~\eqref{eq:classical} (gray bars), except for the additional weak emission at the location of the molecule in the experiment, which could be caused by a non-perfect alignment of the molecule with the center of hemisphere $m=8$ and, also, to the disorder in the energy at different sites.

\begin{figure}[t]
\includegraphics[width=\columnwidth]{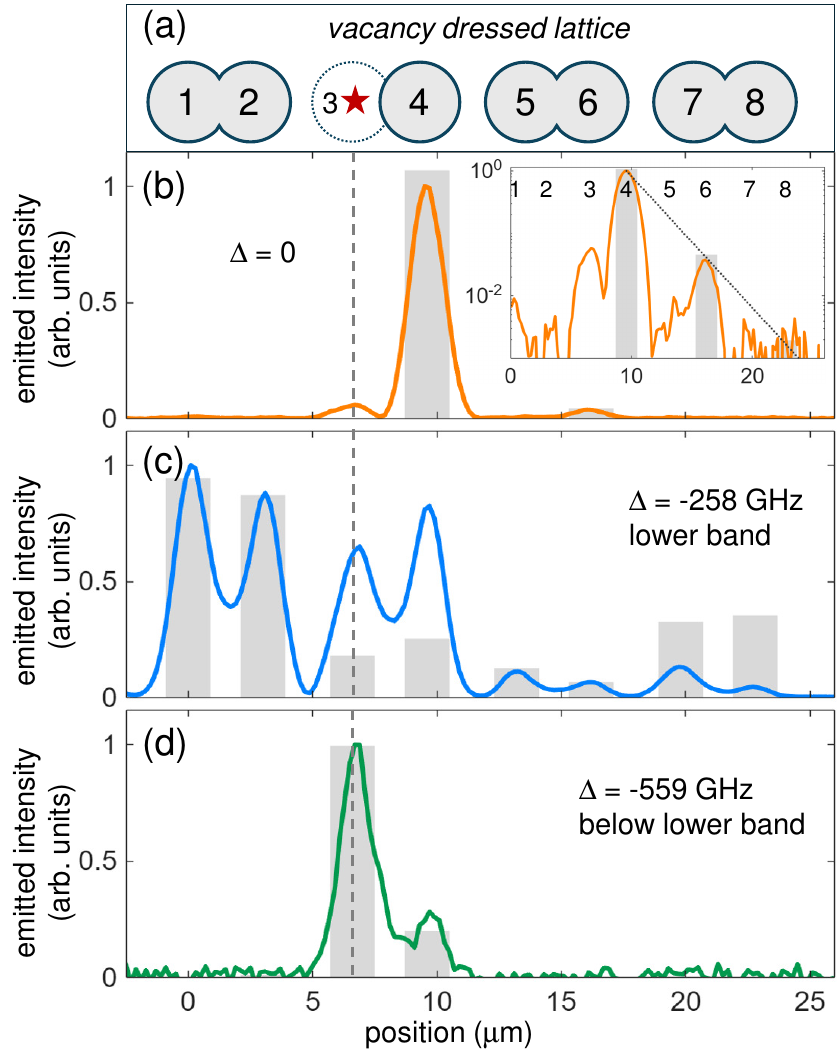}
\caption{\label{fig:chiral_right} 
Directional vacancy-like dressed state with molecule in hemisphere $m=3$. (a) Scheme of the vacancy dressed photonic lattice at the energy of the emitter-photon bound state at $\Delta=0$.
(b)-(d) Solid lines display the measured photoluminescence at the transition energy of a molecule located at site $m=3$ (indicated by a red star and vertical dashes lines) for different emitter-hemisphere detunings.
The grey bars are the steady-state solutions of Eq.~\eqref{eq:classical} at the corresponding detunings.
The inset of (b) displays the data in semi-log scale.
The dotted line shows the expected exponential decay of a topological edge state.
}
\end{figure}

The directionality of the vacancy-like dressed state can be reversed if the molecule is located at an odd site.
Figure~\ref{fig:chiral_right} shows the case of the molecule positioned at site $m=3$.
The vacancy dressed lattice (Fig.~\ref{fig:chiral_right}(a)) cuts the lattice in two.
To the left, we find a single dimer with no eigenmode at $\Delta = 0$, while to the right, it displays an SSH lattice with a non-trivial termination.
Figure~\ref{fig:chiral_right}(b) shows the emitted intensity at $\Delta = 0$.
The photon emission decays exponentially towards the bulk of the lattice in the even sites $m=4$ and $m=6$, as expected for a topological edge state at $\Delta = 0$.
Note that for the same reasons as in Fig.~\ref{fig:chiral_left}(b), we also observe weak emission at the location of the molecule.

\section{Conclusions and outlook}

Summing up, we introduce an optical lattice-QED platform in which individual DBT molecules embedded in anthracene are coupled to lattices of open optical microcavities. As a first demonstration, we investigate the formation of emitter-photon bound states in coupled-resonator systems. In a photonic dimer, we observe the minimal vacancy-like bound state: when the molecular transition is aligned with the bare cavity resonance, destructive interference between the bonding and antibonding modes suppresses emission from the resonator containing the molecule and localizes the photonic component on the opposite site. We then extend this physics to a one-dimensional SSH lattice, where the emitter induces an in-gap localized state whose spatial profile is governed by the equivalent vacancy dressed topological lattice. Depending on the molecular position, this vacancy-like state is localized either on an isolated site or in a sublattice-polarized profile with a directional decay away from the emitter as in an SSH edge state.

This proof-of-principle experiment establishes open-cavity lattices coupled to organic molecules as a versatile route toward many-emitter lattice QED in the optical regime. Recent demonstrations of cavity-mediated hybridization between several DBT molecules in a single photonic resonator provide an encouraging step in this direction~\cite{nobakht_hybridization_2025}. 
A natural next milestone will be the demonstration of photon-mediated coupling between two distant molecules in a lattice, both in the propagating-band regime, where directional and collective emission can be explored, and in the band-gap regime, where bound states can mediate interactions with controlled range and symmetry. 
More broadly, reducing mechanical broadening, improving the homogeneity of the molecule-cavity interface, and integrating active molecular frequency tuning would move this platform toward a programmable optical setup of many-body lattice-QED Hamiltonians with engineered long-range, directional, and topological interactions.

\begin{acknowledgments}
We thank Daniele de Bernardis for useful discussions.
This work has been co-funded by the European Union via the ERC EmergenTopo, 865151 and QUINTESSEnCE, 101088394, Marie Skłodowska-Curie grant agreement No 101108433, and the QUANTERA project MOLAR (PCI2024-153449). 
Views and opinions expressed are however those of the author(s) only and do not necessarily reflect those of the European Union or the European Research Council. 
Neither the European Union nor the granting authority can be held responsible for them
It had been co-funded by the French government through the Programme Investissement d'Avenir (I-SITE ULNE /ANR-16-IDEX-0004 ULNE) managed by the Agence Nationale de la Recherche, the Labex CEMPI (ANR-11-LABX-0007), the region Hauts-de-France, the CDP C2EMPI (R-CDP-24-004-C2EMPI), as well as the French State under the France-2030 programme, the University of Lille, the Initiative of Excellence of the University of Lille, the European Metropolis of Lille.
AGT acknowledges support from the CSIC Research Platform on Quantum Technologies PTI-001, Spanish project Proyecto PID2024-162384NB-I00 financiado por MICIU/AEI/10.13039/501100011033 y por FEDER, UE, from the QUANTERA project MOLAR with reference PCI2024153449 and funded MICIU/AEI/10.13039/501100011033 and by the European Union, the Programa Fundamentos FBBVA through the grant EIC24-1-17304.
T.N. acknowledges the Lumina Quaeruntur fellowship of the Czech Academy of Sciences.
The Würzburg team gratefully acknowledge financial support by the Free State of Bavaria and the Deutsche Forschungsgemeinschaft (DFG, German Research Foundation) through the Würzburg-Dresden Cluster of Excellence ctd.qmat - Complexity, Topology and Dynamics in Quantum Matter (EXC 2147, project-id 390858490).

\end{acknowledgments}

\textit{Data availability} —- The data that support the findings of this article are openly available~\cite{Data}.

\begin{figure}[t]
\includegraphics[width=\columnwidth]{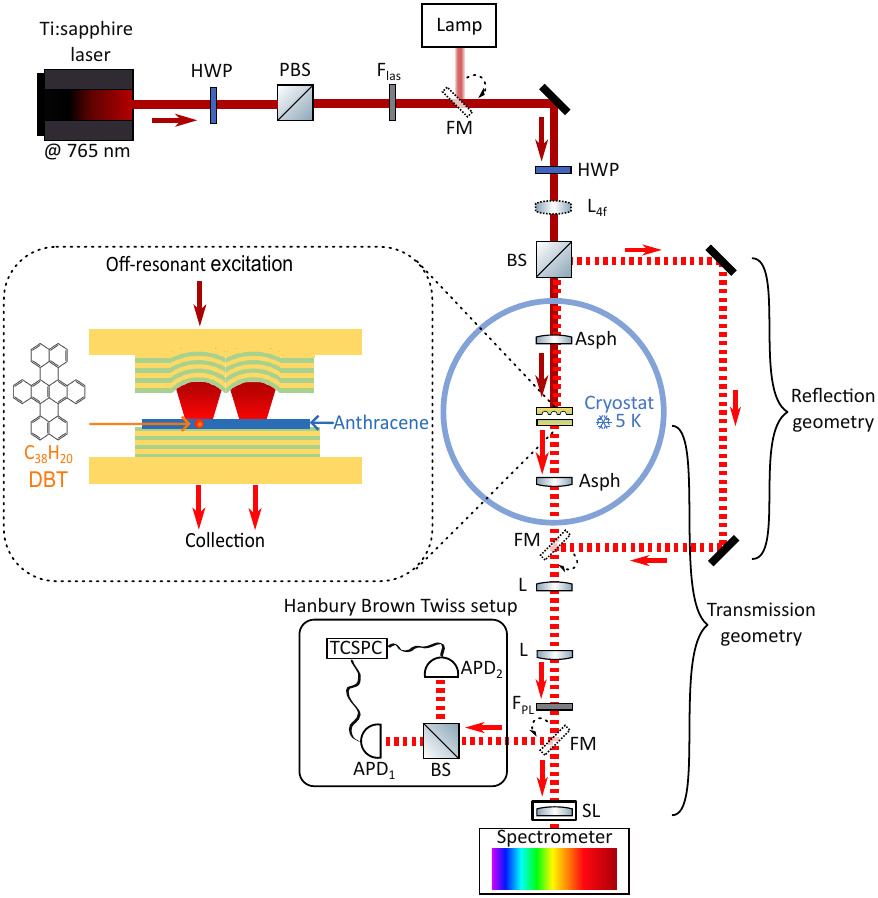}
\caption{\label{fig:setup} 
Experimental setup. HWP, half-waveplate; PBS, polarized beamsplitter; F$_{las}$, laser filter; FM, flip mirror; L$_{4f}$, lens placed at a distance 4f for wide field illumination of the sample; Asph, aspheric lens; BS, beamsplitter; L, lens, F$_{PL}$, PL filter; APD, avalanche photodiode; TCSPC, time-correlated single photon counter; SL, scanning lens.
}
\end{figure}

\appendix

\section{\label{app:set-up} Experimental set-up}

Figure~\ref{fig:setup} displays a sketch of the experimental set-up. A continuous wave Ti:Sapph laser (SolsTiS, M Squared) operating around 765 nm is used to excite the first vibrational level of the DBT molecules placed inside the open cavity. 
Two types of photoluminescence experiments are presented in this article.
The wide field photoluminescence imaged in Fig.~\ref{fig:cavity}(d) is obtained by removing the top mirror and under wide field illumination using L$_{4f}$ in combination with an aspheric lens of focal 8~mm and numerical aperture 0.5 (Thorlabs, A240TM-B), the excitation lens.
The photoluminescence in this case is recovered in reflection geometry and the excitation laser is spectrally filtered out using long-pass filters (Semrock LP785nm RazorEdge and Thorlabs FELH0775).
Using the CCD camera located at the spectrometer, we image the distribution of molecules inside the microcrystal shown in Fig.~\ref{fig:cavity}(d), where each bright point corresponds to the emission from one or several molecules.

All experiments in the cavity configuration are done in transmission geometry (Figs.~\ref{fig:dimer}-\ref{fig:chiral_right}).
The photoluminescence is collected by an aspheric lens of focal 8~mm (Thorlabs, A240TM-B) and sent either to a grating spectrometer (Princeton Instruments) for real space imaging and spectral measurements, or to avalanche photodiodes (Excelitas, SPCM-AQRH-TR) in a Hanbury Brown Twiss configuration connected to a time-tagged single photon counting module (Swabian, Time Tagger Ultra) for intensity correlations measurements.
The signal coming out of the cavity is again spectrally filtered using the above mentioned long pass filters to suppress background signal and transmitted laser light.
The transmission of a broadband incoherent lamp is used to characterize the cavity modes (see, for instance, Fig.~\ref{fig:SSH}(a)).

\begin{figure*}[t]
\includegraphics[width=\textwidth]{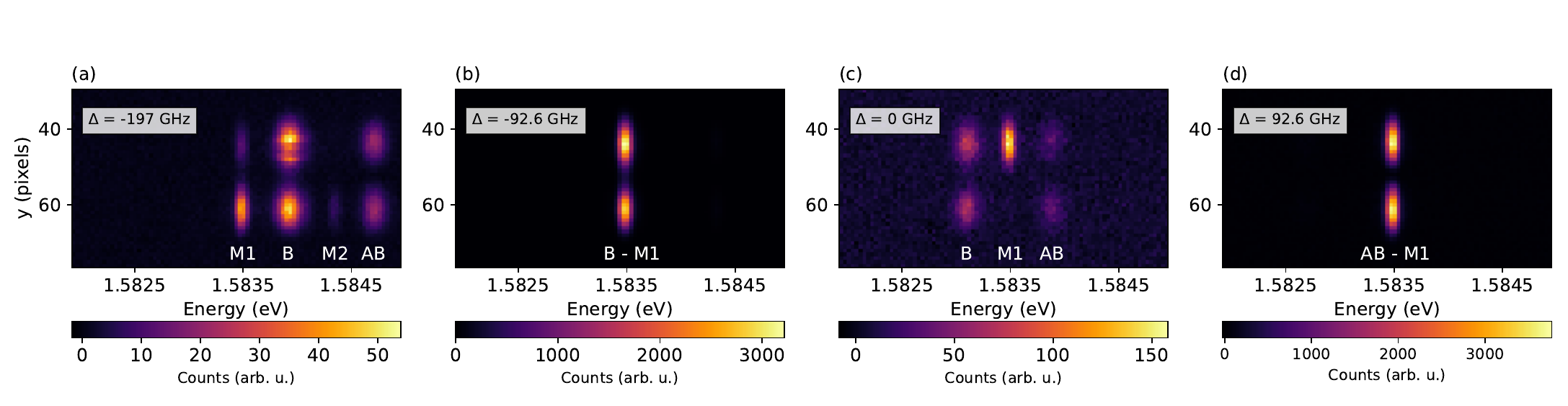}
\caption{\label{fig:spectra} 
Measured photoluminescence spectra for an off-resonant excitation at 765.267~nm in the conditions of Fig.~\ref{fig:dimer} at different emitter-cavity detunings.
The dimer is imaged at the entrance slit plane of the spectrometer, with the slit aligned with the center-to-center axis of the photonic dimer.
The horizontal axis of each image corresponds to the photon energy while the vertical axis $y$ to the vertical spatial dimension of the slit, which in this case is a vertical line going through the center of the hemispheres of the dimer.
For a scale, the center-to-center separation is 3.55~$\mu$m.
The counts are normalized by the laser intensity entering the cavity allowing a direct comparison of the emitted intensities across the different detunings.}
\end{figure*}

The open cavity is placed inside a closed-cycle He cryostat (MyCryofirm) along with the excitation lens and the collection lens, with optical axis being parallel to the optical table. 
In total, there are four stacks of piezoelectric nanopositioners. 
The first stack controls the excitation lens. 
It is composed of a Mechonics piezo ML17 to control the focus, and two Attocube nanopositioners to control the height (ANPz101) and  the horizontal position (ANPx341). 
The latter enables us to completely remove the excitation lens from the optical path when the alignment of the cavity parallelism is being done.
The second stack controls the position and angle of the top mirror with the carved hemispheres.
It is composed of a goniometer (ANGt101) and a rotator (ANR240) to adjust the parallelism, and of a horizontal-axis piezo (ANPx311) and a vertical-axis piezo (ANPz101) .
The long range of the latter allows us to slide the top mirror and expose the bottom mirror while keeping the cryogenic conditions.
The third stack, on top of which is placed the bottom mirror (i.e., the collection side mirror with the anthracene crystals), includes a vertical (ANPz101eXT) and in-plane piezo (ANPx311) to center the molecule of interest, as well as a piezo moving in the optical axis direction (ANPx101) that controls the cavity length.
Finally, the collection lens is placed on three-axis piezo positioners (Mechonics ML17).

\begin{figure}[t]
\includegraphics[width=\columnwidth]{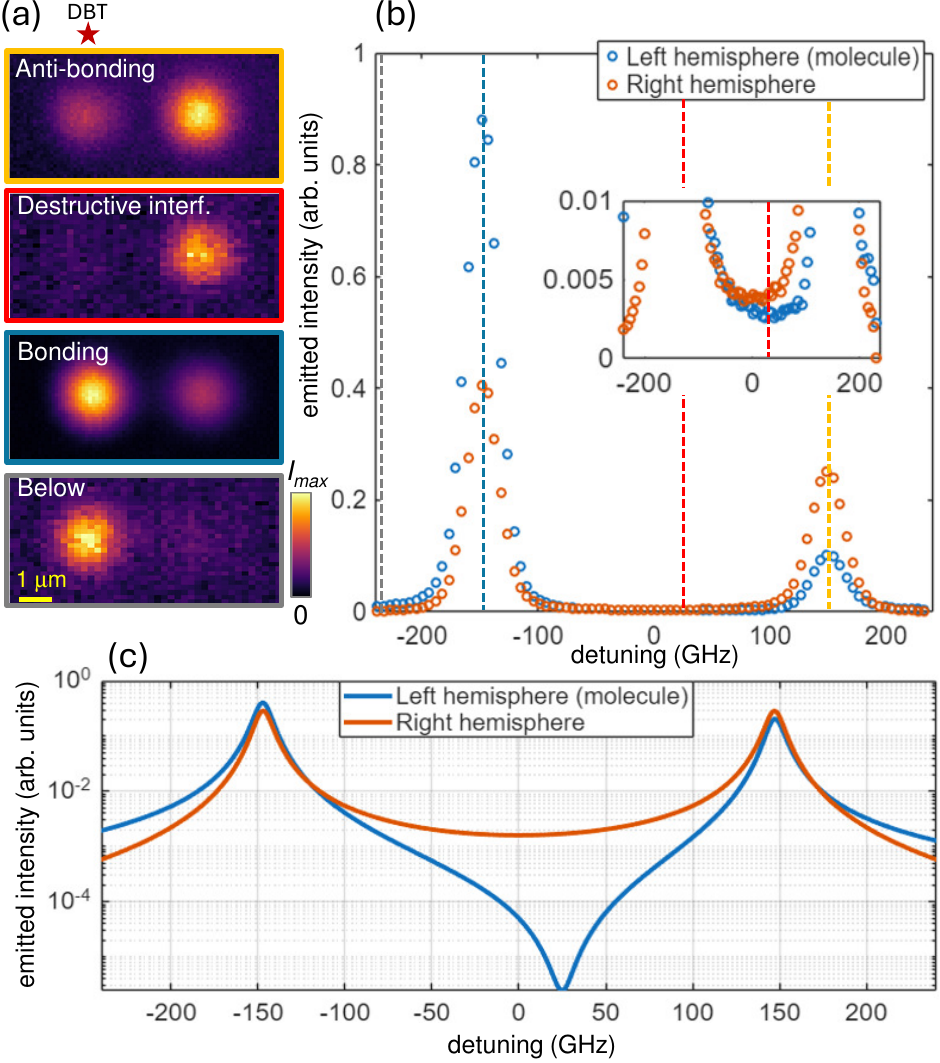}
\caption{\label{fig:dimer_imbalance} 
DBT in an imbalanced dimer of resonators. (a) Spatially resolved photoluminescence of a DBT molecule in a photonic dimer with different on-site energies under off-resonant excitation at different emitter-cavity detunings $\Delta$: the color of the frame corresponds to the detunings marked in dashed lines in (b). 
The molecule is located close to the center of the left hemisphere. 
The destructive interference effect at the hemisphere where the molecule is located takes place at $\Delta=25$~GHz (red frame). 
(b) Emitted intensity from the left (red dots) and right (blue dots) hemispheres as a function of the emitter-cavity detuning.
The inset displays a zoom of the lower intensity section of the graph.
(c) Emitted intensity computed computed from Eq.~\eqref{eq:classical} with $\kappa=11$~GHz, $J=145$~GHz, $E_{left}=E_0-25$~GHz and $E_{right}=E_0+25$~GHz.  }
\end{figure}

The images of the emission of the emitter in the dimer displayed in Figs.~\ref{fig:dimer} and~\ref{fig:dimer_imbalance} are done using standard spectral tomography which spatially and spectrally resolves the photoluminescence of the lattice~\cite{Nardin2009c, Klembt2017, Whittaker2018}: we collect a series of spectrally resolved one-dimensional images in the spectrometer for different horizontal positions of the scanning lens (SL), and we recompose the two-dimensional spatial images at the photon energy corresponding to the emitter's zero-phonon line.

\section{\label{app:convolution} Convolution of cavity transmission with mechanical vibrations}
As shown in Fig.~\ref{fig:dimer}(b), mechanical vibrations of the cavity provide an additional source of broadening in the measured transmission spectra. 
Fluctuations of the cavity length give rise to fluctuations of the cavity resonance frequency. 
As a consequence, the transmission recorded during the acquisition time does not correspond to a single fixed cavity detuning, but rather to an average over a distribution of instantaneous cavity resonance energies.

We account for this effect by assuming that the central energy of the dimer-cavity system fluctuates around its equilibrium value according to
a Gaussian distribution. 
To compute the vibration broadened transmission profiles for a single hemisphere cavity resonance, we do the convolution of a Lorentzian transmission profile with a Gaussian distribution which gives rise to a Voigt profile. 
We apply the same principle to compute emitted intensity of the coupled dimer-cavity system.
The vibration-averaged emission is therefore written as:
\begin{equation}
    I_{vib}(\omega,E_0)=
\frac{\displaystyle \int_{-N\sigma}^{N\sigma} |\psi(\omega,E')|^2G(E',E_0)\,dE'}
{\displaystyle \int_{-N\sigma}^{N\sigma} G(E',E_0)\,dE'},
\end{equation}
where $\psi(\omega,E')$ is the steady-state solution of Eq.~\eqref{eq:classical} for the molecular emission frequency $\omega$ and cavity central energy $E'$. 
The Gaussian weight is defined as $
G(E', E_0)=\exp\left(-4\ln 2\,\frac{(E'-E_0)^2}{\sigma^2}\right)$ with $E_0$ the central energy at equilibrium (in the absence of vibrations), and $\sigma$ the vibration-induced Gaussian width.
In the calculation, the integral is taken from \(-N\sigma\) to \(N\sigma\), with \(N\) large enough to include the relevant part of the Gaussian distribution.

The experimental spectra in Fig.~\ref{fig:dimer} are fitted using a least-squares method applied to \(I_{\mathrm{vib}}\).
This allows us to extract the cavity loss rate \(\kappa = 11\)~GHz and the vibration-induced width \(\sigma=34\)~GHz.

\section{\label{app:data_analysis} Extraction of the photoluminescence intensity at the emission energy of the molecule}

To extract the photoluminescence intensity emitted through the cavity structure, we image the cavity plane onto the entrance slit of a spectrometer in transmission geometry, as shown in Fig.~\ref{fig:setup}.
The slit is aligned and centered following the long axis of symmetry of the specific structure under investigation.
For instance, for the dimer, the slit selects a line passing through the center of both hemispheres.
The photoluminescence is dispersed by the grating of the spectrometer producing two-dimensional images in which the horizontal axis is wavelength (or equivalently, photon energy) and the vertical one is position along the slit.
Figure~\ref{fig:spectra}(a) displays a CCD image of the photoluminescence of a molecule in a dimer in the situation of Fig.~\ref{fig:dimer} for a detuning $\Delta=-197$~GHz, at which the molecular transition is below the bonding mode of the photonic dimer.
The spectrum shows four emission lines.
At $E=1.58347$~eV we observe the direct emission from the molecule (M1).
It is weak and mostly localized at the bottom hemisphere (corresponding to the left hemisphere in Fig.~\ref{fig:dimer}(a)) because it is not in resonance with the bonding or antibonding modes.
Simultaneously, we observe emission from both hemispheres at the energy of the bonding (B) and antibonding (AB) modes, respectively.
The emission of these two lines is at higher energies than the zero-phonon line of the molecule and, therefore, cannot be originated by the phonon and vibration assisted fluorescence from the molecule, which emit at lower energies.
Its origin is most likely the low energy fluorescence from other molecules located within the same dimer with higher zero-phonon line energies, which are also excited by the off-resonant laser excitation employed in the photoluminescence experiments.
Actually, the spectrometer image Fig.~\ref{fig:dimer}(a) also displays the weaker emission from the zero-phonon line of a second molecule labeled M2 at $E=1.58432$~eV.
When the zero-phonon line of molecule M1 is in resonance with the bonding and anti-bonding modes, the emission is completely dominated by these modes, see Fig.~\ref{fig:spectra}(b) and~(d).
Figure.~\ref{fig:spectra}(c) shows the photoluminescence at $\Delta=0$, at which the emitter-photon bound state is visible at the emission energy of M1.

To extract the molecule's photoluminescence intensity at the left and right hemispheres displayed in Fig.~\ref{fig:dimer}(b), we sum the intensity over a window of 5 pixels in the vertical direction centered at the center of each hemisphere, for the energy that corresponds to the zero-phonon line emission at each studied detuning.

\section{\label{app:dimer_imbalance} Photoluminescence in a dimer with different on-site energies}

Figure~\ref{fig:dimer_imbalance} shows the study of the photoluminescence for a molecule located in the left hemisphere of an imbalanced dimer.
We have used the angular alignment piezoactuators of the experiment to add a tilt in the cavity length in the horizontal direction.
In this situation, the onsite energy of the left and right dimer are estimated to be $E_{left}=E_0-25$~GHz and $E_{right}=E_0+25$~GHz for a coupling strength $J=145$~GHz.
For this experiment we have used a dimer with a center-to-center distance of 3.5~$\mu$m.
Figure~\ref{fig:dimer_imbalance}(c) displays the computed emitted intensity from Eq.~\eqref{eq:classical} without the effect of mechanical vibrations.

%

\end{document}